# Multi-asset and generalised Local Volatility. An efficient implementation.

O. Deloire
odeloire@gmail.com

L. Roth
louis.roth2B@gmail.com

(September 2024)

Abstract : this article presents a generic hybrid numerical method to price a wide range of options on one or several assets, as well as assets with stochastic drift or volatility. In particular for equity and interest rate hybrid with local volatility.

Keywords : local volatility, generalised local volatility, trinomial tree, implicit, explicit, Fokker Plank, monte-carlo, ODgrid, interpolation, Steinman, Akima, Steffen, pde, Heston, Hull and White, stochastic volatility, Arrow Debreu, Keys interpolation, trilinear, convertible, equity hybrid.

Contents :

All data used for numerical calculations are shown in Appendix

# 1. One Factor

## 1.1 Generic case

We consider an asset with the following diffusion :

$$dX(t) = \mu(X(t), t)dt + \sigma(X(t), t)dW(t) \quad (0)$$

We assume that a change of variable has been done and that the "non- path dependent" part of the drift as also been taking care of.

Historically the discretisation of such a process has been done via monte-carlo, trees or pde solving, in order to price financial claims.

Monte-carlo being the most generic method but posing challenges for early exercise features.

Pde solving being the most precise, but sometimes hard to implement.

Tree being the simplest but with many limitations.

Here we suggest an hybrid method inspired by trinomial trees. Such trees have been well studied and have merits when volatility and drift are neither time or state dependent. To overcome these issues several solutions have been put forward involving probabilities and/or time steps adjustments.

All approaches follow the same principle, namely, the tree has to recombine. The reason being that a non-recombining tree leads to the well-known dimensionality curse. But, except for that, there is little wrong with it. A monte-carlo simulation can be viewed as different paths chosen in a non-recombining tree.

**Our approach, the ODgrid, is to "combine" a trinomial tree with a fixed "pde like" grid as follows** :

We start by discretising the X(t) diffusion :

$$Xup(t+dt) = X(t) + \mu(X(t),t)dt + \sigma(X(t),t)\sqrt{\frac{3}{2}\,dt}$$

$$Xmd(t+dt) = X(t) + \mu(X(t),t)dt$$

$$Xdn(t+dt) = X(t) + \mu(X(t),t)dt - \sigma(X(t),t)\sqrt{\frac{3}{2}\,dt}$$

Using equiprobabilities :

$$\text{p} = \text{probaUp} = \text{probaMd} = \text{probaDn} = \frac{1}{3}$$

We can verify that $E[dX(t)]$ and $E[dX(t)^2]$are correctly matched.

Then we use the following backward calculation for options :

$$Opt(t) = \frac{1}{3}\,Df\,\big(Optup(t+dt) + Optmd(t+dt) + Optdn(t+dt)\big)$$

Where Df is the discount factor from t+dt to t

Of course this discretisation is by nature non-recombining. It would be if we had Xup, Xmd and Xdn fall exactly on existing nodes at t+dt.

The ODgrid method consists in interpolating the corresponding values from a known grid i.e. knowing a vector of X(t+dt) and corresponding Opt(t+dt) we would find Optup(t+dt) by interpolating Xup(t+dt).

We just need to use a **RELEVANT INTERPOLATION METHOD**. Here, nearest neighbour would be too crude, linear may not capture enough convexity and simple polynomial would lead to instability. ODgrid relies on the use of a Monotone Cubic interpolation. We have several candidates, amongst which Akima [1], M3-A [2] and Stineman [3] (but also Kruger, Carlson, Steffen…).

In pseudo code :

Assuming an ODgrid of values X[i, t] for i=0 to Nx and its corresponding Opt[i, t] at each time step t=0 to N

```
//payoff at maturity t=N
for i=0 to Nx
    Opt[i, N] = payoff(X[i, N], ...)
Next i

//working backwards
For t=N-1 to 0
    For i=0 to Nx
```

$$\mathrm{Xup} = \mathrm{X}[\mathrm{i},\mathrm{t}] + \mu(\mathrm{X}[\mathrm{i},\mathrm{t}],\mathrm{t})\mathrm{dt} + \sigma(\mathrm{X}[\mathrm{i},\mathrm{t}],\mathrm{t})\sqrt{\frac{3}{2}\,\mathrm{dt}}$$

$$\mathrm{Xmd} = \mathrm{X}[\mathrm{i},\mathrm{t}] + \mu(\mathrm{X}[\mathrm{i},\mathrm{t}],\mathrm{t})\mathrm{dt}$$

$$\mathrm{Xdn} = \mathrm{X}[\mathrm{i},\mathrm{t}] + \mu(\mathrm{X}[\mathrm{i},\mathrm{t}],\mathrm{t})\mathrm{dt} - \sigma(\mathrm{X}[\mathrm{i},\mathrm{t}],\mathrm{t})\sqrt{\frac{3}{2}\,\mathrm{dt}}$$

```
        Optup = interpolated value in Opt[*,t+1] corresponding to Xup in X[*,t+1]
        Same for Optmd and Optd using Xmd and Xdn
```

$$\mathrm{Opt}[\mathrm{i},\mathrm{t}] = \frac{1}{3}\mathrm{Df}\,(\mathrm{Optup} + \mathrm{Optmd} + \mathrm{Optdn})$$

```
    Next i
Next t

//value at root
return Opt[0, 0]
```

This method has quite a few advantages :

- It is very generic and works with various drift and volatility specifications
- It is very simple to implement (constant probabilities)
- It has built-in smoothness due to the interpolation method
- It borrows from the tree method by construction
- It borrows from the implicit and explicit pde solving methods via the grid structure
- It borrows from monte-carlo method since all "discount paths" are equiprobable
- It is versatile since you can vary time steps and grid spacing should you want to densify some areas and/or hit specific time and space points (sampling, strike, barriers…)
- It makes it simpler to look at discrete dividends or jumps.
- It is reasonably fast (more details below)
- It has applications in multi dimensions, both with several assets or with one asset and stochastic drift or volatility
- It allows you to specify various correlation structures (states and time dependant) in multi dimensions
- It allows to run several probability measures as well as going forward and backward on the SAME grid, since, with constant probabilities, the drift gets pick up by interpolating values and only the volatility at each grid point matters.

People might question the validity of using an interpolation method, but is worth remembering that it is used in a lot of scientific fields. When the doctor looks at your MRI, when a general looks at a terrain map, when a climatologist analyses the weather, when you take a picture on your mobile phone… Also a Taylor expansion, which we use every day, looks like a nearest neighbour interpolation with the first term, a linear interpolation with the second, a quadratic with third…

Since the core algorithm is simple, we have to focus on two points :

- Setting up the grid i.e. assuming a given time spacing, how many points do we use and how do we space them at each time step to ensure a good trade-off between speed and accuracy
- Interpolating quickly and correctly

On the later issue, it is worth pointing out that the mentioned interpolation methods rely on the knowledge at each time step of abscises values X, their corresponding ordinates Opt but also the "derivatives" at these points. For speed we note that these "derivatives" only need to be calculated once per step. Also if your ODgrid is equally spaced in X, then finding the interval in which you need to perform the interpolation is a straight forward lookup.

Note, that since, in the grid, you can force any X(i, t) to hit specific values (strike, barriers…), you can also set the "derivatives" at specific points (within limits).

## 1.2 Local Volatility

Here we will give an implementation example based on Local Volatility.

Based on the usual log normal, stock price diffusion :

$$\frac{dS(t)}{S(t)} = \big(r(t) - q(t)\big)dt + \sigma(S(t), t)dW(t) \qquad (2)$$

Throughout this article we always take out the non-path dependent terms, via the a variable change, here :

$$S(t) = F(t)e^{X(t)}$$

Where F(t) is the forward price for maturity t :

$$l\,n\left(\frac{F(t)}{F(0)}\right) = \int_0^t \big(r(u) - q(u)\big)du \qquad (3)$$

*F(0) = S(0) =* spot price

So the diffusion we will work with is :

$$dX(t) = -\frac{1}{2}\,\sigma(X(t), t)^2 dt + \sigma(X(t), t)dW(t) \qquad (4)$$

We start by setting up the parameters of the ODgrid, assuming :

- Time t is equally space by N steps from 0 to maturity T by dt increments. The number of time steps N is a model input and dt = T / N.
- Space X is spaced by Nx(t) intervals from Xdn(t) to Xup(t) by dX increments. By definition we have Nx(t) = ( Xup(t) – Xdn(t) ) / dX

In a trinomial tree Nx(t) = 2 * t + 1 , in a grid Nx(t) = constant, in the ODgrid we make Nx(t) time dependent as a parameter used for speed/accuracy trade-off. Note that because of our interpolation method, we need to have Nx(1) greater than 5.

We now detail how we calculate Xdn(t), Xup(t) and dX.

At each time step, we have to make sure that we cover a large enough spectrum of X space.

In a simple Black-Scholes world (constant volatility), we would use a +/- 4 standard deviation :

$$\mathrm{Xdn(t)} = -4\mathrm{BSvol}\sqrt{t}$$

$$\mathrm{Xup(t)} = +4\mathrm{BSvol}\sqrt{t}$$

And as defined in (1) :

$$\mathrm{dX} \;=\; \mathrm{BSvol}\,\sqrt{\frac{3}{2}\mathrm{dt}}$$

In the case of local volatility (strike and time dependant), our arbitrary choice has been to use a +/-4 “standard deviation” based on a volatility reflecting the skew :

$$\mathrm{Xdn(t)} = -4\mathrm{BSvol(Kdn,t)}\sqrt{t}$$

where :

$$Kdn = F(0)\,e^{-4BSvol(F(0),t)\sqrt{t}}$$

and BSvol(K,t) is the BS market volatility for a European option with strike K and maturity t.

Xup(t) set in a similar way with +4 “standard deviation”.

We choose to set dX the increment for X as follows :

$$dX = BSvol(T)\,\sqrt{\frac{3}{2}\mathrm{dt}}\,GridFiness$$

where GridFiness is a model parameter between 0.0 and 1.0.

When GridFiness equals 1.0 and volatility is constant ODgrid is just a trinomial tree, and when GridFiness goes to 0.0 we “converge” to a continuous spacing in X. In practice we use values ranging from  to 1.0 to 0.25.

Alternatives choices for setting up the ODgrid can be made , but the core method remains the same. Assumptions of equal spacing in time and space can be easily relaxed.

A similar approach can be followed in the case of a generic diffusion, depending on the volatility function and assumed terminal distribution for X at each time step.

In pseudo code, putting together the grid set up and backward calculation, the ODgrid algorithm is :

```
//we set up the grid
dt = T / N
dX = BSVol(T) * sqrt(3/2*dt) * GridFiness

For t=0 to N
    Kdn =F(0) * exp( -4 * BSvol(F(0), t*dt) * sqrt(t*dt))
    Xdn(t) = -4 * BSvol(Kdn, t*dt) * sqrt(t*dt)
    Xup(t) = same calculation with +4
    Nx(t) = ( Xup(t) – Xdn(t) ) /dX
Next  t

//payoff at maturity t=N
For i=0 to Nx(N)
    X(i, N) = Xdn(N) + i * dX
    S(i, N) = F(N) * exp(X(i,N)))
    Opt(i,N) = function of S(i,N) e.g Max(0.0 ; S(i,N) - Strike)
Next i

//work backward
```

```
For t=N-1 to 0
    Calc vectSlopes from vectors X(*, t+1) and Opt(*, t+1) for interpolation
    For i=0 to Nx(t)
            X(i, t) = Xdn(t) + i * dX
            S(i, t) = F(t) * exp(X(i,t)))
            Vol = local Volatility for strike S(i,t) and maturity t*dt
            Drift = - Vol * Vol / 2
            //
            XnextUp = X(i, t) + Drift * dt + vol * sqrt(3/2*dt)
            XnextMd = X(i, t) + Drift * dt
            XnextDn = X(i, t) + Drift * dt - vol * sqrt(3/2*dt)
            //
            iup = ( XnextUp – Xdn(t+1) ) / dX
            Optup = interpolate(XnextUp, iup,  X(*, t+1), Opt(*, t+1), vectSlopes)
            //
            imd and idn= same using XnextMd and XnextDn
            Optmd  and Optdn = same using imd and idn
            //
            Opt(i, t) = Df * ( Optup + Optmd + Optdn ) / 3
    Next i
Next t

// value at root
Return Opt(0,0)
```

The following table shows (for several strikes and maturities) the difference between the market implied volatility (using asset1 parametrisation as per Appendix) and the volatility derived from the option premium calculated with the ODgrid. We use a GridFiness of 0.50 and 100 time steps.

| | 0,25 | 0,50 | 1 | 2 | 3 | 5 |
|---|---|---|---|---|---|---|
| 50 | | | | -0,02% | -0,01% | -0,01% |
| 60 | | | -0,01% | 0,00% | 0,00% | 0,00% |
| 70 | | 0,01% | 0,01% | 0,02% | 0,02% | 0,01% |
| 80 | 0,04% | 0,03% | 0,03% | 0,03% | 0,02% | 0,02% |
| 90 | 0,05% | 0,04% | 0,03% | 0,03% | 0,02% | 0,02% |
| 100 | 0,04% | 0,03% | 0,03% | 0,03% | 0,02% | 0,01% |
| 110 | 0,05% | 0,05% | 0,04% | 0,04% | 0,03% | 0,02% |
| 120 | 0,00% | 0,05% | 0,05% | 0,05% | 0,03% | 0,03% |
| 130 | | -0,01% | 0,04% | 0,05% | 0,04% | 0,03% |
| 140 | | | 0,00% | 0,06% | 0,05% | 0,04% |
| 150 | | | | 0,04% | 0,05% | 0,04% |
| 200 | | | | | -0,04% | 0,02% |

## 1.3 Explicit method and Fokker-Plank

Our method could also be used going forward in the grid. For example if we consider the Fokker-Plank equation for the same generic diffusion as before (0).

We have the equation for the probability density (dropping some indexing for reading clarity) :

$$\frac{\partial P}{\partial t} = -\frac{\partial(\mu P)}{\partial X} + \frac{1}{2}\frac{\partial^2(\sigma^2 P)}{\partial X^2} \qquad (5)$$

In the case of non-stochastic rates, this leads directly to the Arrow-Debreu (AD) prices.

We need to use a rectangular ODgrid and have the following explicit discretisation :

Pup(t) at Xup(t) = Xmd(t) + h

Pmd(t) at Xmd(t) = Xmd(t+dt)

Pdn(t) at Xdn(t) = Xmd – h

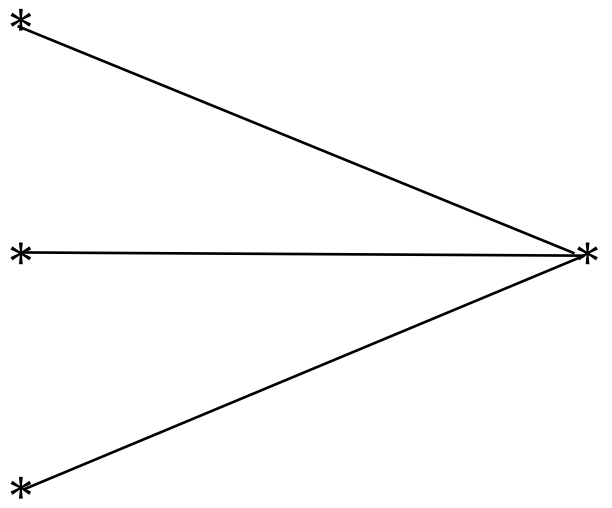

Pmd(t+dt) at Xmd(t+dt)

Since the grid is rectangular with dX spacing, Xmd(t+dt)  and Xmd(t) correspond to the same X point on the ODgrid, but Xup(t) and Xdn(t) do not fall on the grid since they are spaced by h. We would need a before to interpolate the corresponding P values.

Discretising (5) at t with the usual Euler scheme we get :

Pmd(t+dt) = qup Pup(t) + qmd Pmd(t) + qdn Pdn(t)

Where qup, qmd and qdn are coefficients function of h, dt, X, drift and vol.

We have

$$qmd = 1 - \frac{\sigma(Xmd(t), t)^2 dt}{h^2}$$

And for convergence we choose to set :

$$qmd = \frac{1}{3}$$

which implies :

$$h = \sigma(Xmd(t), t)\sqrt{\frac{3}{2}dt}$$

Then we solve for qup and qdn :

$$qup = \frac{1}{3}\left(1 - h\frac{\mu(Xup(t),t)}{\sigma(Xup(t),t)^2}\right)\left(\frac{\sigma(Xup(t),t)}{\sigma(Xmd(t),t)}\right)^2$$

$$qdn = \frac{1}{3}\left(1 + h\frac{\mu(Xdn(t),t)}{\sigma(Xdn(t),t)^2}\right)\left(\frac{\sigma(Xdn(t),t)}{\sigma(Xmd(t),t)}\right)^2$$

In the case of local volatility this simplifies to :

$$qup = \frac{1}{3}\left(1 + \frac{h}{2}\right)\left(\frac{\sigma(Xup(t),t)}{\sigma(Xmd(t),t)}\right)^2$$

$$qdn = \frac{1}{3}\left(1 - \frac{h}{2}\right)\left(\frac{\sigma(Xdn(t),t)}{\sigma(Xmd(t),t)}\right)^2$$

Of course, since we go forward, we would need a boundary condition at t=0, which in this case is a Dirac. However, we might want to start from "some first dt" and use the fact that in first approximation X(dt) follows a normal distribution since :

$$X(dt) \;\cong\; X(0) + \mu(X(0),0)dt + \sigma(X(0),0)\sqrt{dt}\,Z$$

Where X(0), drift(X(0),0) and vol(X(0),0) are all known and Z is N(0,1)

In the case of local volatility, in one dimension, the method described here is of little use, since we can get the AD prices more directly.

With a generic diffusion, it has the advantage of allowing us to go forward in the ODgrid, calculation AD prices and then going backwards on the same grid (for calibration when using a specific volatility function with no closed form solution).

With stochastic rates, in two dimensions, it does bring the ability to calculate generalised local volatility adjustments efficiently going forward, and then use them when pricing going backwards on the same grid.

# 2. Multi Factor cases

## 2.1 Two Factor

Here we look at the case of two “similar assets” i.e. if we price an option based on the Max(S1; S2) it should not matter with asset is defined as S1 and which as S2, this is in contrast with having assets of intrinsically different nature and behaviour, like an asset with stochastic volatility or rate (we will address these in the next sections).

As before, after the relevant variable change we consider :

$$dX1(t) = \mu 1(X1(t), t)dt + \sigma 1(X1(t), t)dW1(t)$$
$$dX2(t) = \mu 2(X2(t), t)dt + \sigma 2(X2(t), t)dW2(t) \qquad (6)$$

$dW1dW2 = \rho_{12}$

Like in (1) with one asset we propose the following discretisation :

( X1(t), X2(t) ) would diffuse in the following states with equal probabilities of 1/5

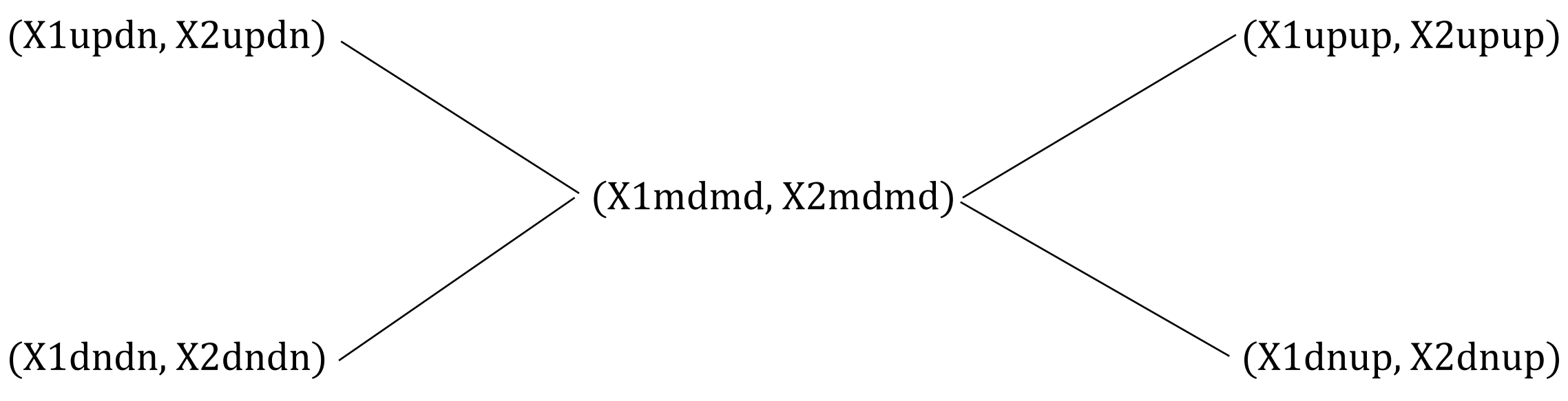

With (for simpler reading, we omit μ and σ dependency on X and t) :

$$\mathrm{X1mdmd} = \mathrm{X1(t)} + \mu 1 \mathrm{dt}$$

$$\mathrm{X2mdmd} = \mathrm{X2(t)} + \mu 2 \mathrm{dt}$$

$$\mathrm{X1upup} = \mathrm{X1(t)} + \mu 1 \mathrm{dt} + \sigma 1 \sqrt{\frac{5}{4}(1+\rho_{12})dt}$$

$$\mathrm{X2upup} = \mathrm{X2(t)} + \mu 2 \mathrm{dt} + \sigma 2 \sqrt{\frac{5}{4}(1+\rho_{12})dt}$$

$$\mathrm{X1updn} = \mathrm{X1(t)} + \mu 1 \mathrm{dt} + \sigma 1 \sqrt{\frac{5}{4}(1-\rho_{12})dt}$$

$$\mathrm{X2updn} = \mathrm{X2(t)} + \mu 2 \mathrm{dt} - \sigma 2 \sqrt{\frac{5}{4}(1-\rho_{12})dt}$$

$$\mathrm{X1dnup} = \mathrm{X1(t)} + \mu 1 \mathrm{dt} - \sigma 1 \sqrt{\frac{5}{4}(1-\rho_{12})dt}$$

$$\mathrm{X2dnup} = \mathrm{X2(t)} + \mu 2 \mathrm{dt} + \sigma 2 \sqrt{\frac{5}{4}(1-\rho_{12})dt}$$

$$\mathrm{X1dndn} = \mathrm{X1(t)} + \mu 1 \mathrm{dt} - \sigma 1 \sqrt{\frac{5}{4}(1+\rho_{12})dt}$$

$$\mathrm{X2dndn} = \mathrm{X2(t)} + \mu 2 \mathrm{dt} - \sigma \sqrt{\frac{5}{4}(1+\rho_{12})dt}$$

And the backward calculation :

$$\mathrm{Opt(t)} = \frac{1}{5} \mathrm{Df}\ (\mathrm{Optmdmd} + \mathrm{Optupup} + \mathrm{Optupdn} + \mathrm{Optdnup} + \mathrm{Optdndn})$$

With the respective “Opt” values being interpolated, this time from a (X1,X2) grid, from the next time step.

The OD grid set up is similar to the 1D, we have just chosen to slightly adjust the calculation of dX1 and dX2 :

$$dX1 = BSvol1(T)\sqrt{\frac{5}{4}(1+|\rho_{12}|)dt}\ GridFiness1$$

$$dX2 = BSvol2(T)\sqrt{\frac{5}{4}(1+|\rho_{12}|)dt}\ GridFiness2$$

And we will have Nx1(t), X1up(t), X1dn(t)….calculated accordingly.

The important part is how you perform the 2D interpolation. And, in order to preserve the symmetry argument we mentioned, we cannot simply interpolate one way and then the other way with the same method used in 1D.

The cleanest method is to use a true bicubic interpolation [4]. This is based on a 2D cubic polynomial so it requires the calculation of 4*4 coefficients. These are based on the first and cross derivatives at each (X1,X2) point of the grid (which only need to be calculated once per time step as before). A 16*16 matrix multiplication needs to be performed, as detailed in reference [4]. This has to be done for each interpolated value (which there are 5 at each node), so it is a rather calculation intensive method, but it has all the desired continuity and smoothness properties.

Another method is to use an interpolation based on **Keys cubic convolution method** [5]. It is a lot faster and easier to implement and has some good properties in terms on continuity and smoothness. The drawback is that it can lead to some unwanted negative values (that would need to be zeroed out, without prejudice).

It is worth pointing out that, like in any numerical method, it comes down to a speed / accuracy trade-off. The GridFiness parameters come into play and if your computer power is limited, Keys method with a few more points is quite fast and accurate.

The following table shows premiums for different call options calculated via Monte Carlo and ODgrid. We use local volatilities for asset1 and asset2 as per the Appendix. Maturity is one year, strikes 80, 100 and 120. We use a GridFiness of 0.5 and 12 time steps. “Basket” correspond to a 50/50 basket call option. “Best” is a bestOf call option. “Spread” is a spread call with strike = K – 100.

| | Mcarlo | ODgrid | Mcarlo | ODgrid | Mcarlo | ODgrid |
|---|---|---|---|---|---|---|
| | basket | basket | best | best | spread | spread |
| 80 | 21,70 | 21,69 | 29,83 | 29,82 | 2,24 | 2,20 |
| 100 | 7,76 | 7,77 | 13,39 | 13,38 | 9,15 | 9,13 |
| 120 | 1,33 | 1,33 | 3,48 | 3,43 | 22,61 | 22,61 |

## 2.2 Three Factor

We start to reach the limits of the method since we now use a 9 points discretisation scheme (i.e. a cube plus its centre) with equal probabilities of 1/9.

Using similar notations as before (6), we have now X1, X2, X3 with correlations r12, r13, r23

Unlike the 2D case, we use a Cholesky decomposition whose coefficients are :

a = r12

$$b = \sqrt{1 - \mathrm{r12}^2}$$

c = r13

$$d = \frac{r23 - r13\, r12}{b}$$

$$e = \sqrt{1 - r13^2 - d^2}$$

We give here the different increment calculation for each of the 9 points :

| | | |
|---|---|---|
| μ1dt | μ2dt | μ3dt |
| μ1dt + $\Sigma_1$ | μ2dt + (a+b) $\Sigma_2$ | μ3dt + (c+d+e) $\Sigma_3$ |
| μ1dt + $\Sigma_1$ | μ2dt + (a+b) $\Sigma_2$ | μ3dt + (c+d−e) $\Sigma_3$ |
| μ1dt + $\Sigma_1$ | μ2dt + (a−b) $\Sigma_2$ | μ3dt + (c−d+e) $\Sigma_3$ |
| μ1dt + $\Sigma_1$ | μ2dt + (a−b) $\Sigma_2$ | μ3dt + (c−d−e) $\Sigma_3$ |
| μ1dt − $\Sigma_1$ | μ2dt − (a−b) $\Sigma_2$ | μ3dt − (c−d−e) $\Sigma_3$ |
| μ1dt − $\Sigma_1$ | μ2dt − (a−b) $\Sigma_2$ | μ3dt − (c−d+e) $\Sigma_3$ |
| μ1dt − $\Sigma_1$ | μ2dt − (a+b) $\Sigma_2$ | μ3dt − (c+d−e) $\Sigma_3$ |
| μ1dt − $\Sigma_1$ | μ2dt − (a+b) $\Sigma_2$ | μ3dt − (c+d+e) $\Sigma_3$ |

where :

$$\Sigma_i = \sigma_i \sqrt{\frac{9}{8} dt}$$

The setting up of the grid and backward interpolation / calculation is as before, only here, due to resources constraints, we have used a simple **trilinear interpolation** [6] for each points.

The following table shows premiums for different call options calculated via Monte Carlo and ODgrid. We use local volatilities for asset1, asset2 and asset3 as per the Appendix. Maturity is one year, strikes 80, 100 and 120. We use a GridFiness of 0.20 and 12 steps. “Basket” correspond to an equi-weighted basket call option. “Best” is a bestOf call option.

| | Mcarlo | ODgrid | Mcarlo | ODgrid |
|---|---|---|---|---|
| | basket | basket | best | best |
| 80 | 21,65 | 21,70 | 35,88 | 35,98 |
| 100 | 8,20 | 8,22 | 18,70 | 18,75 |
| 120 | 1,95 | 1,95 | 7,37 | 7,35 |

## 2.3 Higher dimensions

It is clear that the dimensionality curse is hard to avoid (even more so with our interpolation method), and that a monte-carlo method is going to be hard to beat in this case.

But, for further research and intellectual curiosity, I would mention here an avenue that could be investigated.

In higher dimensions, in order to avoid the "Nx1*Nx2*Nx3*Nx4…." curse resulting from using a rectangular grid, we would need to fill the space using a "quasi random numbers" method (Korobov good lattice points [7] comes to mind). In this case we would control in a linear way the number of points used in the discretisation.

Then, we would need to find the kth nearest neighbours to the point we want to interpolate. By doing so we may lose a lot of the computational advantage previously gained, but there are a lot of very interesting publications from other "science fields" dealing with this subject that may provide us with a solution (also using a quasi-random sequence may help).

Finally, we would ned to use a higher dimensional method of interpolation (inverse distance weighting, kernel…). Again, people outside of finance have looked at that.

Not having gone down that path, I cannot comment beyond this initial though.

# 3. Hybrid cases

## 3.1 Stochastic rates

We start by considering a simple equity / stochastic rates model (using Hull White [8]) :

$$\frac{dS(t)}{S(t)} = r(t)dt + \sigma_S dW_S$$

$$dr(t) = \big(\theta(t) - kr(t)\big)dt + \sigma_r dW_r$$

With : lnS(t) = X1(t) and r(t) = φ(t) + X2(t)

we have :

$$dX1(t) = \left(r(t) - \frac{1}{2}{\sigma_S}^2\right)dt + \sigma_S dW_S$$

$$dX2(t) = -kX2(t)dt + \sigma_r dW_r$$

$$dW_S dW_r = \rho_{Sr}$$

k is the mean reversion coefficient for rates

rates and equity volatilities, $\sigma_S$and $\sigma_r$, are assumed constant

φ (t) is a non-stochastic drift to match the market yield curve

$$\varphi(\mathrm{t}) = \mathrm{f}(0,\mathrm{t}) + \frac{{\sigma_r}^2}{2k^2}\left(1 - e^{-kt}\right)^2$$

f(0,t) is the market “instantaneous” forward rate

$\theta$ () ,function of φ (), k and $\sigma_r$is not explicitly needed

Having this parametrisation, we know we can price europeen equity options just by using a volatility adjustment, which will easily allow us to check numerical results.

The adjustment is as follow :

$$BSrateAdjVol(t)^2 = \sigma_S{}^2 + \frac{2}{k}\sigma_S\sigma_r\rho_{Sr}\left(1 - \frac{1-e^{-kt}}{kt}\right) + \frac{\sigma_r{}^2}{k^2}\left(1 + \frac{1-e^{-2kt}}{2kt} - 2\frac{1-e^{-kt}}{kt}\right)$$

This volatility can then be used in a plain BS option pricing formula.

This simple example allows us to see how we make the ODgrid method adapt to different diffusions in two dimensions.

First when setting up the grid, we can use the fact that we know the standard deviation of the rate distribution in the Hull and White model :

$$stdDev(t) = \sigma_r\sqrt{\frac{1-e^{-2kt}}{2k}}$$

So our X2up and X2dnwould be set a +/- 4 times stdDevR(t)

The rest of the ODgrid set up similar to section 2.1

Secondly and more importantly, when interpolating on the grid we can take in consideration the nature of the processes we are dealing with. In the previous two assets cases we noted that it was important to take into account the "symmetric" role of each asset.

Here it is quite different. We will price different options on an equity linked claim, but at the same time we want to take into account the stochastic nature of the drift (rates). So in this case, we can differentiate our interpolation methods (same for the case of stochastic volatility).

So we choose to use a **monotonic cubic interpolation for the equity and a linear for the rates**. This helps in term of calculation/speed but also make sense, since the rates process is mean reverting (like stochastic volatility).

The core algorithm is similar to the one used in section 2.1, with the relevant grid set up and the following backward calculation in pseudo code :

$$r(t) = X2(t) + \varphi(t)$$

$$X1mdmd = X1(t) + \left(r(t)\ - \frac{{\sigma_S}^2}{2}\right)dt$$

$$X2mdmd = X2(t) - kX2(t)dt$$

$$X1upup = X1(t) + \left(r(t)\ - \frac{{\sigma_S}^2}{2}\right)dt + \sigma_S\sqrt{\frac{5}{4}(1+\rho_{Sr})dt}$$

$$X2upup = X2(t) - kX2(t)dt + \sigma_r\sqrt{\frac{5}{4}(1+\rho_{Sr})dt}$$

$$X1updn = X1(t) + \left(r(t)\ - \frac{{\sigma_S}^2}{2}\right)dt + \sigma_S\sqrt{\frac{5}{4}(1\ -\ \rho_{Sr})dt}$$

$$X2updn = X2(t) - kX2(t)dt - \sigma_r\sqrt{\frac{5}{4}(1-\rho_{Sr})dt}$$

$$X1dnup = X1(t) + \left(r(t)\ - \frac{{\sigma_S}^2}{2}\right)dt - \sigma_S\sqrt{\frac{5}{4}(1-\rho_{Sr})dt}$$

$$X2dnup = X2(t) - kX2(t)dt + \sigma_r\sqrt{\frac{5}{4}(1-\rho_{Sr})dt}$$

$$X1dndn = X1(t) + \left(r(t)\ - \frac{{\sigma_S}^2}{2}\right)dt - \sigma_S\sqrt{\frac{5}{4}(1+\rho_{Sr})dt}$$

$$X2dndn = X2(t) - kX2(t)dt - \sigma_r\sqrt{\frac{5}{4}(1+\rho_{Sr})dt}$$

Here for the backward calculation we have :

$$\mathrm{Opt(t)} = \frac{1}{5}e^{-r(t)dt}(\mathrm{Optmdmd} + \mathrm{Optupup} + \mathrm{Optupdn} + \mathrm{Optdnup} + \mathrm{Optdndn})$$

With Optupup linearly interpolated in “X2” between two values calculated by monotonic cubic interpolation in “X1”. The same for the other Opt values.

The following table shows (for several strikes and maturities) the difference between the $BSrateAdjVol$ market implied volatility (calculated from formula above with a 20% equity volatility and HW rate parametrisation as per Appendix) and the volatility corresponding to the option premium calculated with the ODgrid. We use a GridFiness of 0.50 for equity, 0.50 for rates and 50 time steps.

| | 0,50 | 1,00 | 3,00 |
|---|---|---|---|
| 90,00 | 0,03% | 0,04% | 0,02% |
| 100,00 | 0,07% | 0,07% | 0,06% |
| 110,00 | 0,05% | 0,05% | 0,09% |

## 3.2 Stochastic volatility

We deal here with volatility models like Litpon [9], Blacher [10] and Heston [11].

We choose Heston for convenience. But the former models would be dealt with similarly, with maybe the use of a two dimensional forward calculation of Arrow-Debreu prices for ease of calibration, as mentioned in 1.3 and implemented in 3.3.

After the usual variable change, we start with the following diffusions :

$$dX1(t) = -\frac{v(t)}{2}dt + \sqrt{v(t)}dWs$$

$$dv(t) = k\big(v1 - v(t)\big)dt + \sigma\sqrt{v(t)}dWv \quad (8)$$

dWs dWv $= \rho_{Sv}$

We use the same methodology as in 2.1 and 3.1 by interpolating cubically in X1 and then linearly in v.

The following table shows premiums for one year call options calculated via Monte Carlo and ODgrid. We use Heston parameters as per the Appendix. We use a GridFiness of 1.0 for equity, 1.0 for volatility and 50 time steps.

| | Mcarlo | ODgrid |
|---|---|---|
| 90,00 | 12,89 | 12,85 |
| 100,00 | 6,63 | 6,60 |
| 110,00 | 2,75 | 2,77 |

## 3.3 Generalised Local Volatility

Similar to section 3.1 we assume the following diffusions :

$$dX1(t) = \left(r(t) - \frac{\Sigma adj_S(X1,t)^2}{2}\right) dt + \Sigma adj_S(X1,t) dWs$$

$$dX2(t) = -kX2(t)dt + \sigma_r dWr$$

Except that, in this case, as shown in [12], [13], [14] and [15] we have to use an adjusted version Σadj of the local volatility Σ to take into account the stochastic nature of interest rates.

Following the results from the previous references we have :

$$\Sigma^2(X1,t) = \frac{Nume}{Deno}$$

$$Nume = Vega\ \left(\frac{\sigma}{2t} + \frac{d\sigma}{dt} + f(0,t)\ K\ \frac{d\sigma}{dK}\right)$$

$$Vega = K\ Df(t)\ \sqrt{t}\ n(d2)$$

$$\Sigma adj(X1,t)^2 = \frac{Nume + AdjFactor}{Deno}$$

Or :

$$\Sigma adj(X1,t)^2 = \varSigma^2(X1,t) + \frac{AdjFactor}{Deno} = \varSigma^2(X1,t)\ \left(1 + \frac{AdjFactor}{Nume}\right) \qquad (9)$$

Where :

$$AdjFactor = -K\ \mathbb{E}[r(t)\ D(X1,X2,t)\ \mathbb{1}(S_t > K)]\ -\ Vega\, K\ f(0,t)\frac{d\sigma}{dK} + K\ Df(t)\ f(0,t)\ N(d2)$$

N(), n(), d1, d2 and σ are the usual BS variables

The Expectation is taken under the **risk neutral measure**, so that :

$$\mathbb{E}[r(t)\ D(X1,X2,t)\ \mathbb{1}(S_t > K)] = \sum_{X1,X2} r(t)\ AD(X1,X2,t)\ \mathbb{1}(S_t > K)$$

Note that the heavy side function is better defined with a value of 0.5 at zero.

The calculation of these expectations can be done in an easy additive way on the grid, starting from the highest values of $S_t$ and working your way down to the lowest.

**We use our method in two ways, one going forward in the ODgrid to calculate the AD prices using the Fokker-Plank equation to obtain the adjusted local volatility at each point, and then, to go backward using the same grid, with the exact same method as in 3.1.**

Since the backward calculation is similar to what has been done before (which is the point of the ODgrid), we focus on the former part of the calculation.

In two dimensions, the Fokker-Plank equation writes (dropping indexation for clarity) :

$$\frac{dAD}{dt} = -r(t)AD + k\frac{d(x2AD)}{dx2} - \frac{d\left(\left(r(t) - \frac{\Sigma adj_S{}^2}{2}\right)AD\right)}{dx1} + \frac{1}{2}\frac{d^2\left(\Sigma adj_S{}^2 AD\right)}{dx1^2} + \frac{1}{2}\sigma_r{}^2\frac{d^2 AD}{dx2^2}$$
$$+\rho_{Sr}\sigma_r\frac{d^2(\Sigma adj_S AD)}{dx1dx2}$$

Here the extra term r(t)AD is a result of going from the probability density to the AD prices.

Now we use the usual Euler scheme to discretize the above equation on our **rectangular** fixed ODgrid (spaced by dX1 and dX2 chosen for the backward calculations), using a **different dx1 and dx2** (at each point), and regrouping the different terms we have :

$$AD(t+1,i,j) = \sum_{k=-1}^{1}\sum_{l=-1}^{1} q(t,k,l)AD(t,i+k,j+l) \qquad (10)$$

Where, AD(t, i+k, J+l) correspond to the state values X1(i,j) + k dx1 and X2(i,j) + l dx2

With the first term corresponding to the central node (by construction, the same for t and t+1) :

$$\mathrm{q(t,i,j)} = 1.0 - \left(\mathrm{X2(i,j)} + \varphi(\mathrm{t}) + \frac{\sigma_r{}^2}{dx1^2} + \frac{\Sigma adj_S(t,i,j)^2}{dx2^2}\right)dt$$

Like we did in section 1.3 we decide to choose dx1 and dx2 so that q(t, i, j) is close to 1/9 :

$$dx1 = \Sigma adj_S(t,i,j)\sqrt{\frac{9}{4}dt}$$

and

$$dx2 = \sigma_r\sqrt{\frac{9}{4}dt}$$

All these values are known are time t, since here we are calculating AD(t+1,i , j).

With this choice of dx1 and dx2, we can calculate all the other coefficients q(t, , ) and AD(t, , ) needed by interpolation as we have done before.

Having calculated our AD(t+1, i, j) from (10) we can know get the $\Sigma adj_S$ for that time step using (9).

Having calculated the $\Sigma adj_S$ for that step we can now calculate our AD for the following ...and so on by recurrence till maturity.

Having calculated all $\Sigma adj_S$ for all nodes in our ODgrid we just go backward using our previously described method (see 3.1).

It is worth mentioning that, like in any Fokker-Plank implementation, we should start at t=0 with a Dirac i.e. AD(0,0,0) = 1. But it may be a better idea to start from "some t=dt" step and use the fact that in first approximation :

$$X1(dt) \;\cong\; X1(0) + \left(f(0,0) - \frac{\sigma_S(X1(0),0)^2}{2}\right)dt + \sigma_S(X1(0),0)\sqrt{dt}\,Z1$$

$$X12(dt) \cong X2(0) - kX2(0)dt + \sigma_r\sqrt{dt}Z2$$

Where all parameters are known for t=0 and Z1, Z2 are joint bi-normal biN(0, 1, $\rho_{Sr}$).

Hence the first step approximation for all AD(1, i, j) is straight forward.

The volatility adjustment factor calculation (via the AdjFactor) should be dealt with carefully since we are mixing continuous and discrete time "equations" i.e. taking the ratio of two (very) small terms, one from a discretization and one from an analytical calculation, can lead to numerical problems, especially when you get to the wings (cutoff may be necessary, and a good way of doing it is by looking at the "Vega" value in the equations, which makes sense since we are dealing with a volatility adjustment).

The following table shows (for several strikes and maturities) the difference between the market implied volatility (using asset1 parametrisation as per Appendix) and the one derived from the option premium calculated with the Odgrid (using HW rates parametrisation as per Appendix). We use a GridFiness of 0.33 for equity, 0.50 for rates and weekly time steps.

We can see that we have a decent calibration (quick and non-iterative), for a grid that can then be used to price hybrid / convertible products.

| | 1,00 | 2,00 | 3,00 | 5,00 |
|---|---|---|---|---|
| 70,00 | 0,06% | 0,08% | 0,07% | 0,06% |
| 80,00 | 0,06% | 0,05% | 0,05% | 0,04% |
| 90,00 | 0,05% | 0,04% | 0,05% | 0,04% |
| 100,00 | 0,06% | 0,04% | 0,04% | 0,03% |
| 110,00 | 0,05% | 0,04% | 0,04% | 0,04% |
| 120,00 | 0,04% | 0,02% | 0,03% | 0,03% |
| 130,00 | 0,03% | 0,03% | 0,02% | 0,03% |

# 4. Appendix

Here is the data framework used for calculating numerical examples in this article.

We always assume zero dividends.

Except for part 3.1 and 4 , we assume zero rates.

In these sections, we model the market zero coupon interest rate curve as :

$$R(t) = R1 + (R0 - R1)\frac{1 - e^{-ct}}{ct}$$

R0 is the short rate, R1 the long rate and c is a curvature parameter

We use R0 = 2% , R1 =4%   and c = 1.0

We model the market BS implied volatility as an SVI type :

$$BSvol(K,t)^2 = \frac{w}{2t}\left(1 + rx\varphi(w) + \sqrt{1 - r^2 + \left(r + x\varphi(w)\right)^2}\right)$$

With :

$$x = ln\left(\frac{K}{Spot}\right)$$

$$vol(t) = v1\sqrt{1 + \left(\frac{1 - e^{-ct}}{ct}\right)\left(\frac{v0^2}{v1^2} - 1\right)}$$

$$w = vol(t)^2\, t$$

$$\varphi(w) = \frac{a}{w^b\,(1 + w)^{(1-b)}}$$

v0, v1 and c are atm volatility parameters (short vol, long vol and curvature)

r, a and b are skew parameters (r for curvature , a and b to match 1Y and 5Y 95/105 skew)

We use v0 = v1 = 25% , c = 5 and r = 0.8 , a = -0.718 , b = 0.424 for the first asset X1

We use v0 = v1 = 20%, c = 5 and = 0.8, a = -0.299, b = 0.451 for the second asset X2

We use v0 = v1 = 30%, c = 5 and = 0.8, a = 0.0, b = 0.392 for the third asset X3

The relevant local volatilities are calculated using the usual methods.

For HW parameters we use :

k=0.05 for mean reversion

$\sigma_r$= 2% for rate volatility

$\rho_{Sr} = -0.3$ for equity/rate correlation

For Heston we use :

v(0) = 2.9% for initial variance

v(inf) = 2.9% for variance mean reversion level

σ = 35% for variance volatility

k = 3.0 for mean reversion coefficient

ρ = - 0.5 for equity/variance correlation